\documentclass[journal]{IEEEtran}

\usepackage{graphicx,graphics,subfigure}
\usepackage{amsfonts,bm}
\usepackage{amsmath}
\usepackage{dcolumn}
\usepackage{bigstrut}
\usepackage{pdfpages}
\usepackage{epstopdf}
\usepackage{color}
\usepackage{booktabs}
\usepackage{url}
\usepackage{multirow}
\usepackage{multicol}
\usepackage{textcomp}
\usepackage{algorithm}
\usepackage{algorithmic}

\graphicspath{{Figure/}}
\usepackage[normalem]{ulem}
\useunder{\uline}{\ul}{}
\title{Deep Convolutional Sparse Coding Networks for Image Fusion}

\begin{document}
	\author{Shuang Xu, Zixiang Zhao, Yicheng Wang, Chunxia Zhang, Junmin Liu, \textit{Member IEEE} and Jiangshe Zhang

		\thanks{S.Xu and Z. Zhao are co-first authors. S. Xu, Z. Zhao, Y. Wang, C. Zhang, J. Liu, and J. Zhang  are with the School of Mathematics and Statistics, Xi'an Jiaotong University, Xi'an 710049, China (e-mail: shuangxu@stu.xjtu.edu.cn; zixiangzhao@stu.xjtu.edu.cn; yc\_wang@stu.xjtu.edu.cn; cxzhang@mail.xjtu.edu.cn; junminliu@mail.xjtu.edu.cn;  jszhang@mail.xjtu.edu.cn). }
		
	}
\maketitle
\begin{abstract}
	Image fusion is a significant problem in many fields including digital photography, computational imaging and remote sensing, to name but a few. Recently, deep learning has emerged as an important tool for image fusion. This paper presents three deep convolutional sparse coding (CSC) networks for three kinds of image fusion tasks (i.e., infrared and visible image fusion, multi-exposure image fusion, and multi-modal image fusion). The CSC model and the iterative shrinkage and thresholding algorithm are generalized into dictionary convolution units. As a result, all hyper-parameters are learned from data. Our extensive experiments and comprehensive comparisons reveal the superiority of the proposed networks with regard to quantitative evaluation and visual inspection.  
\end{abstract}
\section{Introduction}
Image fusion is a fundamental topic in image processing \cite{image_fusion}, and its aim is to generate a fusion image by combining the complementary information of source images \cite{Liu_review}. This technique has been applied to many scenarios. For example, in military, infrared and visible image fusion (IVF) is helpful for object detection and recognition \cite{Ma_review}. In digital photography, high dynamic range (HDR) imaging can be solved by multi-exposure image fusion (MEF) to generate high-contrast and informative images \cite{MEF-SSIM2018TCI}.  

Over the past a few decades, numerous image fusion algorithms have been proposed, where transform based algorithms are very popular \cite{Liu_review}. They transform source images into feature domain, detect the active levels, blend the features and at last apply the inverse transformer in order to obtain the fused image. Recently, deep neural networks have emerged as an effective tool in image fusion \cite{Liu_review}. They are divided into three groups: (1) Autoencoder based methods. This is a deep learning variant of transform based algorithms. The transformers and inverse transformers are replaced by encoders and decoders, respectively  \cite{li2018densefuse}. (2) Supervised methods. For multi-focus image fusion, there are ground truth images in the synthetic datasets \cite{Liu2017_MFF}. For MEF, Cai et al. constructed a large dataset providing the reference images by comparing 13 MEF/HDR algorithms \cite{SICE}. Owing to the strong fitting ability, supervised learning networks are suitable for these tasks. (3) Human visual system based methods. In the case without reference image, by taking prior knowledge into account and setting proper loss functions, researchers designed regression \cite{UMFF,MEF-NET} or adversarial \cite{ma2019fusiongan} networks to make fusion images satisfy human visual systems. However, it is found that many algorithms are evaluated on a limited number of cherry-picked images. Thus, their generalizations still remain unknown. It leaves room for possible improvement with reasonable and interpretable formulations. 

Convolutional sparse coding (CSC) has been successfully applied to computer vision tasks on account of its high performance and robustness \cite{Multi-CSC,CSC-RGB-NIR}. The CSC model is generally solved by the iterative shrinkage and thresholding algorithm (ISTA), but the results significantly depend on hyper-parameters. To address this problem, the CSC model and ISTA are generalized into some dictionary convolutional units (DCUs) which are put in the hidden layers of neural networks. In this manner, the hyper-parameters (e.g. penalty parameters, dictionary filters and thresholding functions) in DCUs are learnable. Based on the novel unit, we design deep CSC networks for three fusion tasks, including IVF, MEF, and multi-modal image fusion (MMF). In our experiments, we employ relatively large test datasets to make a comprehensive and convincing evaluation. Experimental results show that the deep CSC networks outperform the state-of-the-art (SOTA) methods in terms of both objective metrics and visual inspection. Besides, our networks are with high reproducibility. The remainder of this paper is organized as follows. Section \ref{sec:DCU} converts the CSC and ISTA into a DCU. Then, in section \ref{sec:net} we design three DCU based networks for IVF, MEF and MRF tasks. The extensive experiments are reported in section \ref{sec:experiment}. Section \ref{sec:conclusion} concludes this paper. 

\section{Dictionary Convolutional Units}\label{sec:DCU}
In dictionary learning, CSC is a typical method for image processing. Given an image $\bm{x}\in R^{c\times h\times w}$ ($c=1$ for gray images and $c=3$ for RGB images) and $q$ convolutional filters $\bm{d}\in R^{q\times c\times s\times s}$, CSC can be formulated as the following problem:
\begin{equation}
\min_{\bm{z}} \frac{1}{2}\|\bm{x}-\bm{d}*\bm{z}\|_2^2+\lambda g(\bm{z}),
\end{equation}
where $\lambda$ is a hyperparameter, $*$ denotes the convolution operator, $\bm{z}\in R^{q\times h\times w}$ is the sparse feature map (or say, code) and $g(\cdot)$ is a sparse regularizer. This problem can be solved by ISTA, and it is easy to write the updating rule for feature maps as below,
\begin{equation}\label{eq:csc_update}
\bm{z}^{(k+1)} \leftarrow \mathrm{prox}_{\lambda/\rho}\left(\bm{z}^{(k)}+\frac{1}{\rho}\bm{d}^T*(\bm{x}-\bm{d}*\bm{z}^{(k)})\right),
\end{equation}
where $\rho$ is the step size and $\bm{d}^T\in R^{c\times q\times s\times s}$ is the flipped version of $\bm{d}$ along horizontal and vertical directions. Note that $\mathrm{prox}(\cdot)$ is the proximal operator of the regularizer $g(\cdot)$. If $g(\cdot)$ is the $\ell_1$-norm, its corresponding proximal operator is the soft shrinkage thresholding (SST) function defined by
$
\mathrm{SST}_{\gamma}(x) = \mathrm{sign}(x)\mathrm{ReLU}(|x|-\gamma),
$
where $\mathrm{ReLU}(x)=\max(x,0)$ is the rectified linear unit and $\mathrm{sign}(x)$ is the sign function. CSC provides a pipeline to extract features of an image, but its performance highly depends on the configuration of $\{\lambda,\rho,\bm{d}\}$. By the principle of algorithm unrolling \cite{Algorithm_Unrolling,LCSC,LSC}, the ISTA of CSC can be generalized as a unit in neural networks. We employ two convolutional units, $\mathrm{Conv}_i (i=0,1)$, to replace $\bm{d}$ and $\bm{d}^T/\rho$, and proximal operator $\mathrm{prox}(\cdot)$ is extended to the activation function $f(\cdot)$. Hence, Eq.(\ref{eq:csc_update}) can be rewritten as
\begin{equation}\label{eq:dict_conv}
\bm{z}^{(k+1)} = f\left( {\rm BN}\left( \bm{z}^{(k)}+\mathrm{Conv}_1(\bm{x}-\mathrm{Conv}_0(\bm{z}^{(k)})) \right) \right),
\end{equation}
where we also take batch normalization (BN) into account.
It is worth pointing out that, except for SST, the activation function can be freely set to alternatives (e.g., ReLU, parametric ReLU (PReLU) and so on) if the regularizer $g(\cdot)$ is not set to $\ell_1$-norm. In what follows, Eq. (\ref{eq:dict_conv}) is called a dictionary convolutional unit (DCU). By stacking DCUs, the original CSC model can be represented as a deep CSC neural network. 

In addition, stacking DCUs is interpretable to representation learning. $\mathrm{Conv}_0$ serves as a decoder, since it maps $\bm{z}^{(k)}$ from feature space to image space. And $\mathrm{Conv}_1$ serves as an encoder, since it maps the residual between the original image $\bm{x}$ and the reconstructed image $\mathrm{Conv}_0(\bm{z}^{(k)})$ from image space to feature space. Then, the encoded residual is added to the current code $\bm{z}^{(k)}$ for updating. Eventually, the output passes through BN and an activation function for non-linearity. This process can be regarded as an iterative auto-encoder.

\section{Deep Convolutional Sparse Coding Based Image Fusion}\label{sec:net}
\begin{figure*}
	\centering
		\includegraphics[width=\linewidth]{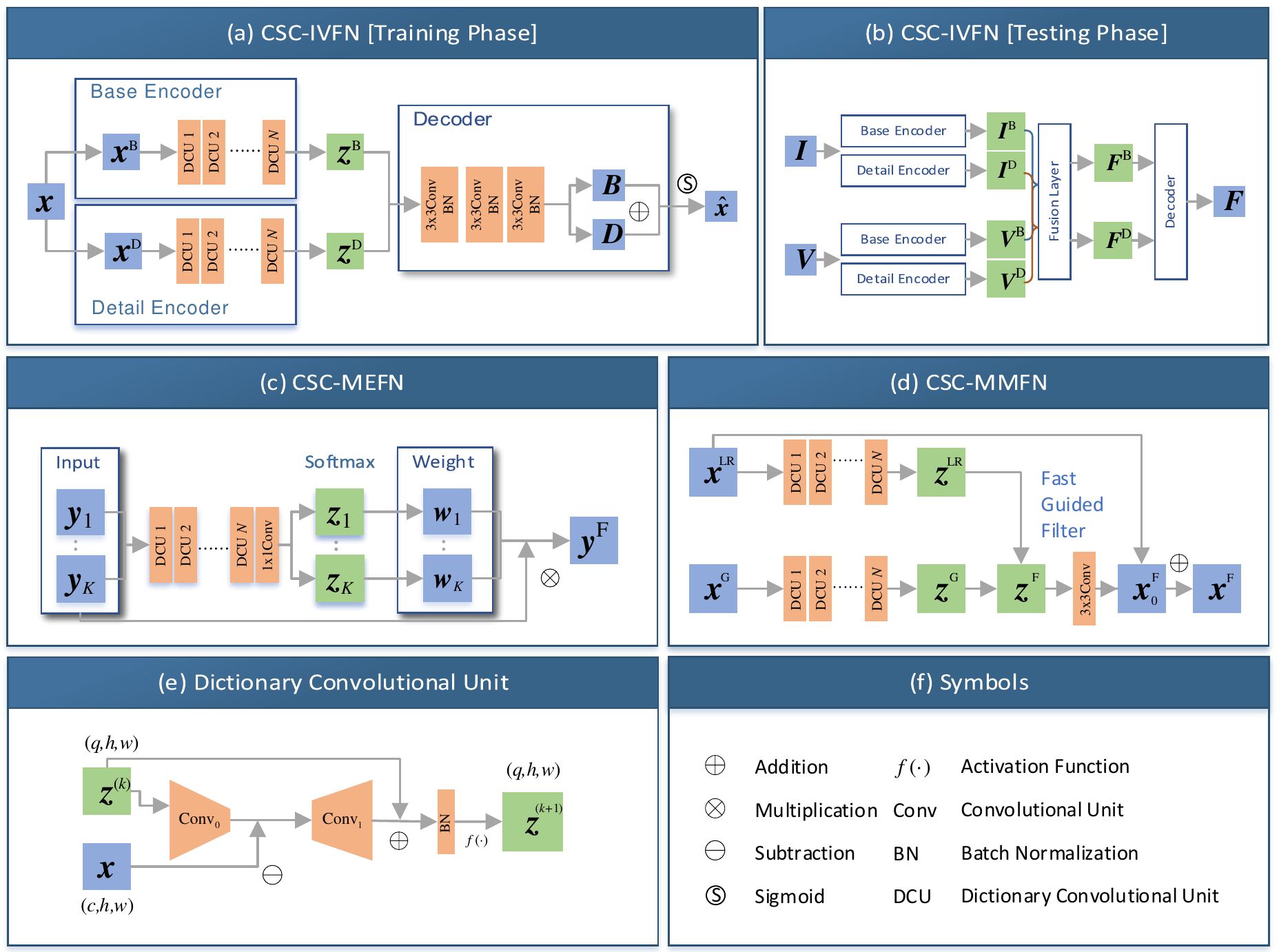}
	\caption{Network structure.}
	\label{fig:Net}
\end{figure*}

In this section, we apply deep CSC neural networks to the image fusion problem, and exhibit three paradigms of model formulation for three different image fusion tasks. 
\subsection{Infrared and Visible Image Fusion}
By combining autoencoders and the CSC model, we propose a CSC-based IVF network (CSC-IVFN), which can be regarded as a flexible data-driven transformer. In the training phase, we train CSC-IVFN in the autoencoder fashion. In the testing phase, features obtained by the encoder of CSC-IVFN are fused and the fusion image is decoded by a decoder. 

\subsubsection{Training Phase}
The architecture is displayed in Fig.~\ref{fig:Net} (a). Firstly, the input image $\bm{x}$ \footnote{In the training phase, both infrared and visible images are indiscriminately denoted by $\bm{x}$.} is decomposed into a base image $\bm{x}^{\rm B}$ containing low-frequency information and a detail image $\bm{x}^{\rm D}$ containing high-frequency textures. Similar to \cite{liu2016image,DBLP:journals/corr/abs-1905-03590}, $\bm{x}^{\rm B}$ is obtained by applying a box-blur filter to $\bm{x}$, and as for the detail image there is $\bm{x}^{\rm D}=\bm{x}-\bm{x}^{\rm B}$. Then, the base and detail images pass through $N$ stacked DCUs, and we will get the final feature maps, that is, $\bm{z}^{\rm D}$ and $\bm{z}^{\rm B}$. And next we feed them into a decoder to decode the base and detail images. Finally, they are combined to reconstruct the input image. Here, the output is activated by a sigmoid function to make sure that the values range from 0 to 1. The loss function is mean squared error (MSE) plus structural similarity (SSIM) loss,
\begin{equation}\label{equ:Loss_total}
L^{\rm IVF}=\frac{1}{hw}\left(\|\bm{x}-\hat{\bm{x}}\|_2^2+\lambda^{\rm IVF} \frac{1-{\rm SSIM}(\bm{x},\hat{\bm{x}})}{2}\right),
\end{equation}
where $\lambda^{\rm IVF}$ is a trade-off parameter to balance the MSE and SSIM \cite{SSIM}. Note that MSE is used to keep the spatial consistency and SSIM guarantees local details in terms of structure, contrast and brightness \cite{SSIM}.

\subsubsection{Testing Phase}
After training a CSC-IVFN, there is a transformer (encoder) and inverse transformer (decoder). In the test phase, CSC-IVFN is feed with a pair of infrared and visible images. In what follows, we use $\bm{I}^{\rm B}$, $\bm{I}^{\rm D}$, $\bm{V}^{\rm B}$ and $\bm{V}^{\rm D}$ to represent the base and detail feature maps of infrared and visible images, respectively.
As exhibited in Fig.~\ref{fig:Net}~(b), a fusion layer is inserted between encoder and decoder in the test phase. It can be expressed by a unified merging operation $\mathcal{F}(\cdot)$, 
\begin{equation}\label{}
\begin{aligned}
\bm{F}^{\rm B}=\mathcal{F}(\bm{I}^{\rm B},\bm{V}^{\rm B})=\bm{w}_1^{\rm B}\otimes \bm{I}^{\rm B}\oplus\bm{w}_2^{\rm B}\otimes \bm{V}^{\rm B}, \\
\bm{F}^{\rm D}=\mathcal{F}(\bm{I}^{\rm D},\bm{V}^{\rm D})=\bm{w}_1^{\rm D}\otimes \bm{I}^{\rm D}\oplus\bm{w}_2^{\rm D}\otimes \bm{V}^{\rm D}.
\end{aligned}
\end{equation}
Here, $\otimes$ and $\oplus$ are element-wise product and addition. There are three popular fusion strategies: 

\begin{enumerate}
	\item Average strategy: $\displaystyle{\bm{w}_1^{\rm B}=\bm{w}_1^{\rm D}=\bm{w}_2^{\rm B}=\bm{w}_2^{\rm D}=0.5}$.
	\item IVF $\ell_1$-norm fusion strategy\cite{li2018densefuse,liu2016image}: It uses the $\ell_1$-norm of patches as the active level. For base weights, there are
	\begin{equation}\label{}
	\begin{aligned}
	\bm{w}_1^{\rm B} &= {\Gamma \left( {{{\left\| \bm{I}^{\rm B} \right\|}_1}} \right)} / \left[{\Gamma \left( {{{\left\| \bm{I}^{\rm B} \right\|}_1}} \right) + \Gamma \left( {{{\left\| \bm{V}^{\rm B} \right\|}_1}} \right)}\right],\\ \bm{w}_2^{\rm B}&=1-\bm{w}_1^{\rm B},
	\end{aligned}
	\end{equation}
	where $\Gamma(\cdot)$ is a $3\times3$ mean filter. The detail weights $\bm{w}_1^{\rm D}$ and $\bm{w}_2^{\rm D}$ can be obtained in the same way.
	\item Saliency-weighted fusion strategy \cite{DBLP:journals/corr/abs-1905-03590}: To highlight and retain the saliency target and information, the fusion weight of this strategy is determined by the saliency degree. We take base weights as an example.
	Firstly, the saliency value of $\bm{I}^{\rm B}$ at the $k$th pixel can be obtained by
	$
	\bm{S}_{\bm{I}}^{\rm B}(k)=\sum_{i=0}^{255} \bm{H}^{\rm B}_{\bm{I}}(i)|\bm{I}^{\rm B}(k)-i|,
	$
	where $\bm{I}^{\rm B}(k)$ is the value of the $k$th pixel and $\bm{H}^{\rm B}_{\bm{I}}(i)$ is the frequency of pixel value $i$. The initial weight at the $k$th pixel is $
	\tilde{\bm{w}}_1^{\rm B}(k) = {\bm{S}_{\bm{I}}^{\rm B}(k)}/\left[{\bm{S}_{\bm{I}}^{\rm B}(k) + \bm{S}_{\bm{V}}^{\rm B}(k)}\right]$ and $\tilde{\bm{w}}_2^{\rm B}(k)=1-\tilde{\bm{w}}_2^{\rm B}(k)$.
	To prevent region boundaries and artifacts, the weight map is refined  via the guided filter $\mathcal{G}(\cdot,\cdot)$ with the guidance of base and detail feature maps:
	\begin{equation}\label{equ:saliency2}
	\begin{aligned}
		\bm{w}_1^{\rm B} &= \mathcal{G}({\tilde{\bm{w}}_1^{\rm B}},\bm{I}^{\rm B})/\left[\mathcal{G}({\tilde{\bm{w}}_1^{\rm B}},\bm{I}^{\rm B}) + \mathcal{G}({\tilde{\bm{w}}_2^{\rm B}},\bm{I}^{\rm V}) \right]
	,\\ \bm{w}_2^{\rm B}&=1-\bm{w}_1^{\rm B}.
	\end{aligned}
	\end{equation}
\end{enumerate}

\subsection{Multi-Exposure Image Fusion}
Most of MEF algorithms fall under the umbrella of weighted summation framework,
$
\bm{f}=\sum_{k=1}^{K} \bm{w}_k\otimes\bm{x}_k,
$
where $\{\bm{x}_k\}_{k=1}^K$ are source images, $\{\bm{w}_k\}_{k=1}^K$ are the corresponding weight maps, $\bm{f}$ is the fused image and $K$ denotes the number of exposures. We propose a CSC-based MEF network (CSC-MEFN). Different from CSC-IVFN, CSC-MEFN is an end-to-end network. Here DCUs extract feature maps, which are then used to predict weight maps to generate the fusion image. To avoid chroma distortion, the proposed CSC-MEFN works in the YCbCr space, and its channels are denoted by $\bm{y}_k,\bm{b}_k$ and $\bm{r}_k$. As shown in Fig. \ref{fig:Net} (c), Y channels $\{\bm{y}_k\}_{k=1}^K$ pass through CSC-MEFN one-by-one. At first, CSC-MEFN stacks $N$ DCUs to code the Y channels. Then, it is followed by a $1\times1$ convolutional unit to get the final code $\bm{z}_k$. Thereafter, the codes $\{\bm{z}_k\}_{k=1}^K$ are converted into weight maps $\{\bm{w}_k\}_{k=1}^K$ by softmax activation. At last, the fused Y channel $\bm{y}^{\rm F}$ is obtained by $\bm{y}^{\rm F}=\sum_{k=1}^{K} \bm{w}_k\otimes\bm{y}_k$. As for the Cb channels, we employ the MEF $\ell_1$-norm fusion strategy, i.e.,
$
\bm{b}^{\rm F} = \sum_{k=1}^{K}\|\bm{b}_k-0.5\|_1\bm{b}_k/\sum_{k=1}^{K}\|\bm{b}_k-0.5\|_1.
$
So Cr channels do. After the separate fusion of three channels, the fusion image $\bm{f}$ is transformed from YCbCr to RGB space. Eventually, we apply a post-processing \cite{l1_l0TM}: the values at 0.5\% and 99.5\% intensity level are mapped to [0,1], and values out of this range are clipped. 

CSC-MEFN is supervised by improved MEFSSIM \cite{MEF-SSIM2018TCI}. It evaluates the similarity between source images $\{\bm{x}_k\}_{k=1}^K$ and the fusion image $\bm{f}$ in terms of illumination, contrast and structure. Our experimental results show that MEFSSIM often leads to haloes. Essentially, halo artifacts result from the pixel fluctuation in the illumination map (i.e., Y channel). To suppress haloes, we propose a halo loss defined by the $\ell_1$-norm on gradients of the illumination map,
$
L_{\rm halo}=\|\nabla \bm{y}^{\rm F}\|_1,
$
where $\nabla$ denotes the image gradient operator (see details in supplementary materials). In our experiments, $\nabla$ is implemented by horizontal and vertical Sobel filters. In summary, given the penalty parameter $\lambda^{\rm MEF}$, the loss function of CSC-MEFN is expressed by
\begin{equation}
L^{\rm MEF} = \frac{1}{hw}\left(-{\rm MEFSSIM}+\lambda^{\rm MEF} L_{\rm halo}\right).
\end{equation}

\subsection{Multi-Modal Image Fusion}
Owing to the limitation of multispectral imaging devices, multispectral images (MS) contain enriched spectral information but with low resolution (LR). One of the promising techniques for acquiring a high resolution (HR) MS is to fuse the LRMS with a guidance image (e.g. panchromatic or RGB images). This problem is a special MMF task. We present a CSC-based MMF network (CSC-MMFN) for the general MMF task. It is assumed that LR and guidance images are represented by
$\bm{x}^{\rm LR} = \bm{d}^{\rm LR}*\bm{z}^{\rm LR}$ and 
$\bm{x}^{\rm G} = \bm{d}^{\rm G}*\bm{z}^{\rm G},$
respectively. Given the dictionary of HR images $\bm{d}^{\rm F}$, the HR image is represented by 
\begin{equation}
	\bm{x}^{\rm F} = \bm{d}^{\rm F}*(\bm{z}^{\rm LR})\uparrow.
\end{equation} 
The symbol $\uparrow$ denotes the upsampling operator. According to this model, CSC-MMFN separately extracts codes of $\bm{x}^{\rm LR}$ and $\bm{x}^{\rm G}$ by two sequences of DCUs, and we utilize the fast guidance filter to super-resolve $\bm{z}^{\rm LR}$ with the guidance of $\bm{z}^{\rm G}$. At last, the HR image is recovered by a $3\times3$ convolutional unit. The loss function is set to MSE between ground-truth and fusion images. 

\section{Experiments}\label{sec:experiment}

\begin{table*}[h]
	\centering
	\caption{Datasets employed in this paper.}
	\resizebox{\textwidth}{!}{
		\begin{tabular}{ccccccccc}
			\toprule
			\multirow{2}[4]{*}{IVF} & Training &       & \multicolumn{2}{c}{Validation} &       & \multicolumn{3}{c}{Test} \\
			\cmidrule{2-2}\cmidrule{4-5}\cmidrule{7-9}      & FLIR-Train &       & NIR-Water & NIR-OldBuilding &       & TNO   & FLIR-Test & NIR-Country \\
			\midrule
			\# Pairs & 180   &       & 51    & 51    &       & 40    & 40    & 52 \\
			Illumination & Day\&Night &       & Day   & Day   &       & Night & Day\&Night & Day \\
			Objectives & Individual\&Stuff &       & Scenery & Building &       & Individual\&Stuff & Individual\&Stuff & Scenery \\
			\bottomrule
		\end{tabular}%
	}
	\resizebox{0.55\textwidth}{!}{
		\begin{tabular}{ccccccc}
			\toprule
			\multirow{2}[4]{*}{MEF} & Training &       & \multicolumn{2}{c}{Validation} &       & Test \\
			\cmidrule{2-2}\cmidrule{4-5}\cmidrule{7-7}      & SICE-Train &       & SICE-Val & TCI2018 &       & HDRPS \\
			\midrule
			\# Pairs & 466   &       & 51    & 24    &       & 44 \\
			\# Exposures & 6-28  &       & 5-20  & 3-30  &       & 9 \\
			\bottomrule
		\end{tabular}%
	}
	\resizebox{0.344\textwidth}{!}{
		\begin{tabular}{cc}
			\toprule
			MMF   & Cave \\
			\midrule
			\# Train/Validation/Test & 22/4/6 \\
			LR Image & Multispectral \\
			Guide Image & RGB \\
			\bottomrule
		\end{tabular}%
	}
	\label{tab:dataset}
\end{table*}

Here we elaborate the implementation and configuration details of our networks. Experiments are conducted to show the performance of our models and the rationality of network structures. For each task, our experiments utilized training, validation and test datasets. The hyperparameters are determined by validation set.  

\subsection{Infrared and Visible Image Fusion}
\begin{figure*}[!]
	\centering
	\subfigure[Infrared]{\includegraphics[width=0.16\textwidth]{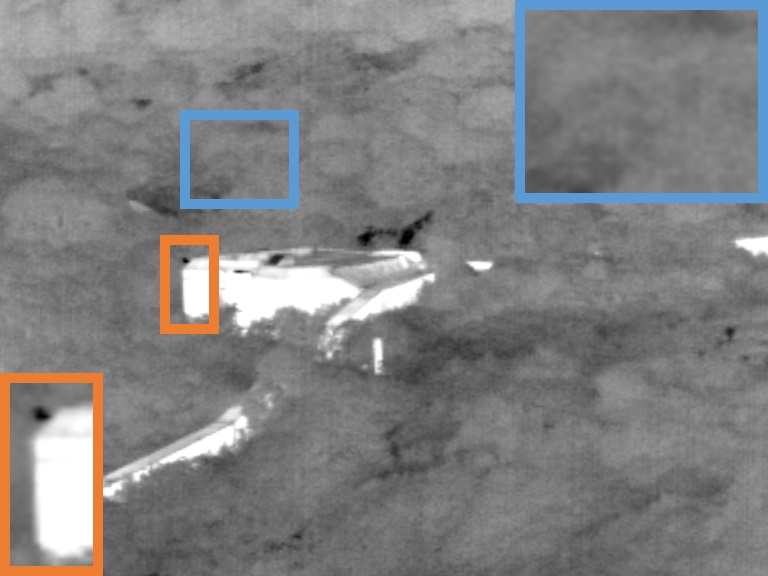}} 
	\subfigure[Visible]{\includegraphics[width=0.16\textwidth]{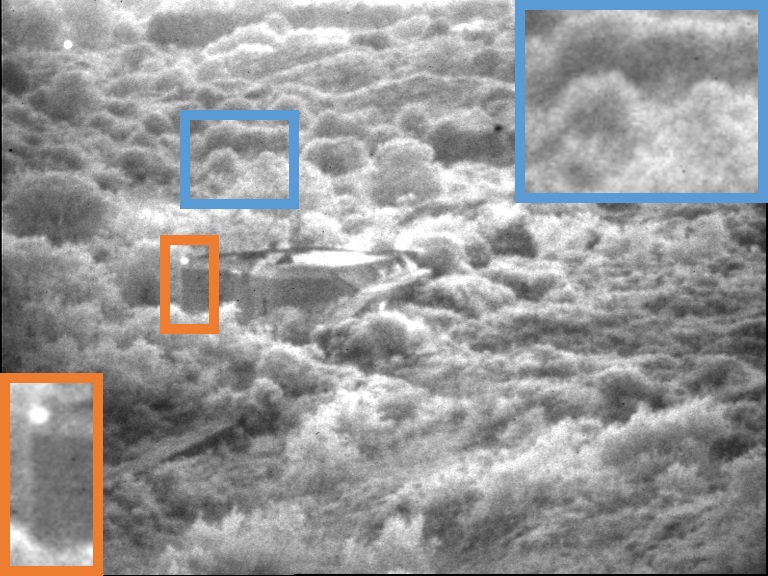}}
	\subfigure[ADKT]{\includegraphics[width=0.16\textwidth]{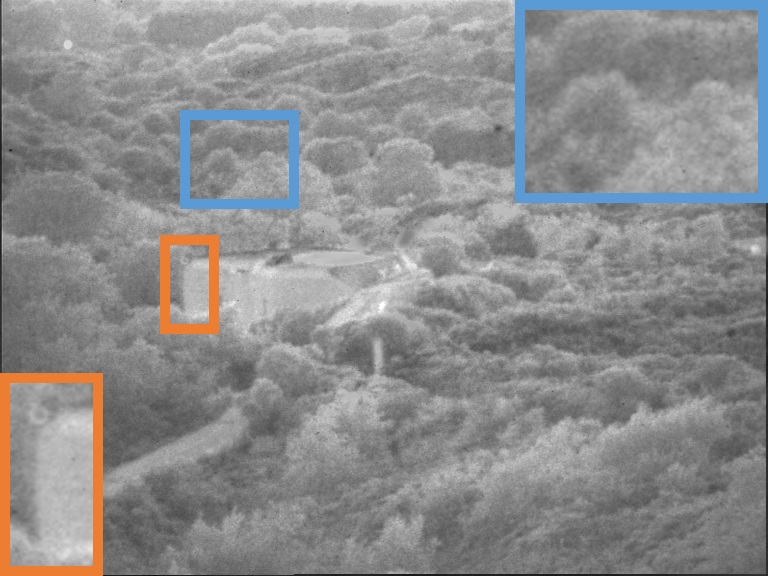}} 
	\subfigure[CSR]{\includegraphics[width=0.16\textwidth]{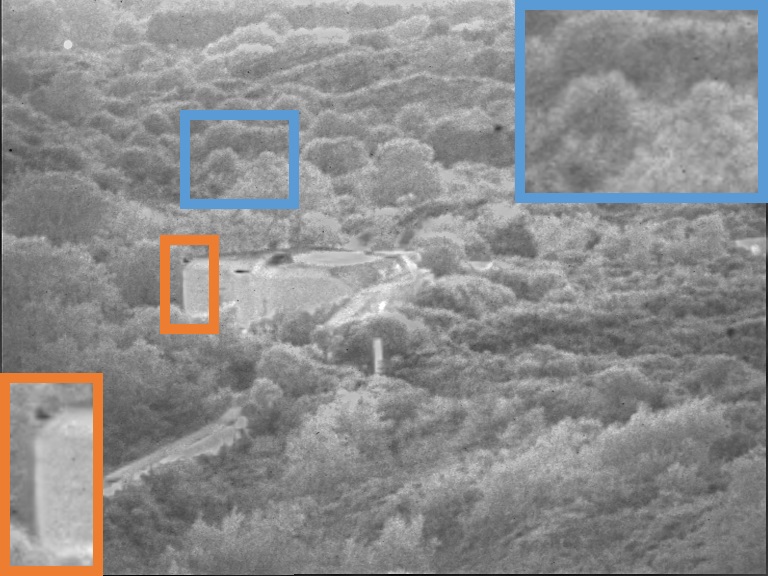}} 
	\subfigure[DeepFuse]{\includegraphics[width=0.16\textwidth]{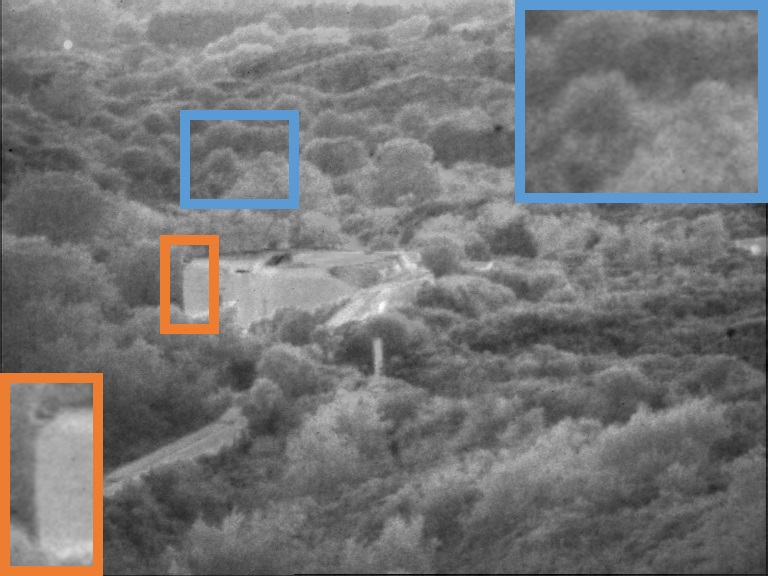}} 
	\subfigure[DenseFuse]{\includegraphics[width=0.16\textwidth]{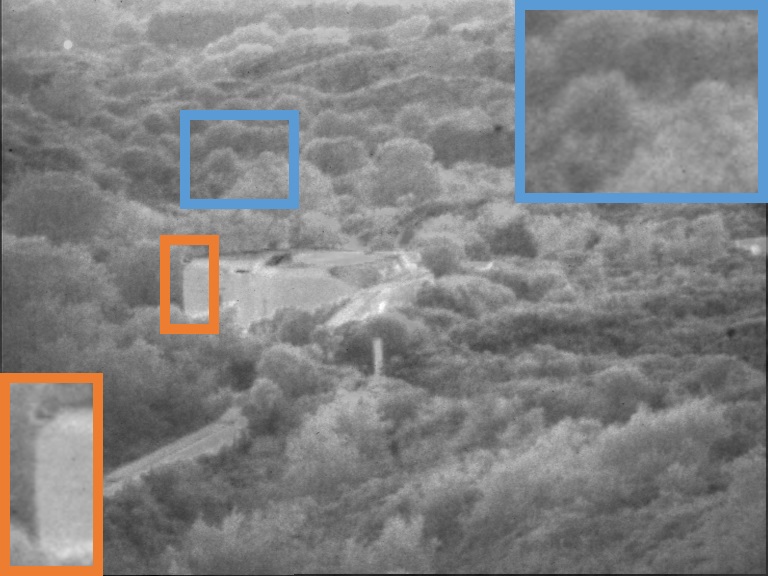}} 
	\subfigure[DLF]{\includegraphics[width=0.16\textwidth]{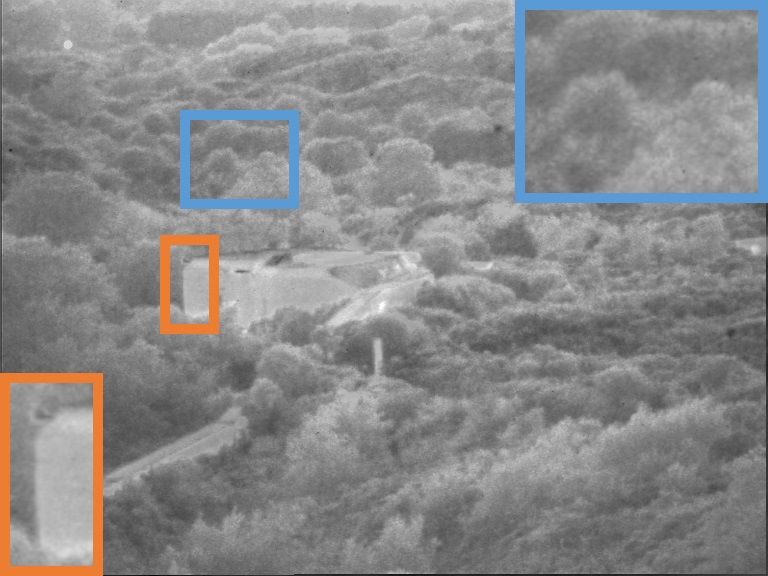}} 
	\subfigure[FEZL]{\includegraphics[width=0.16\textwidth]{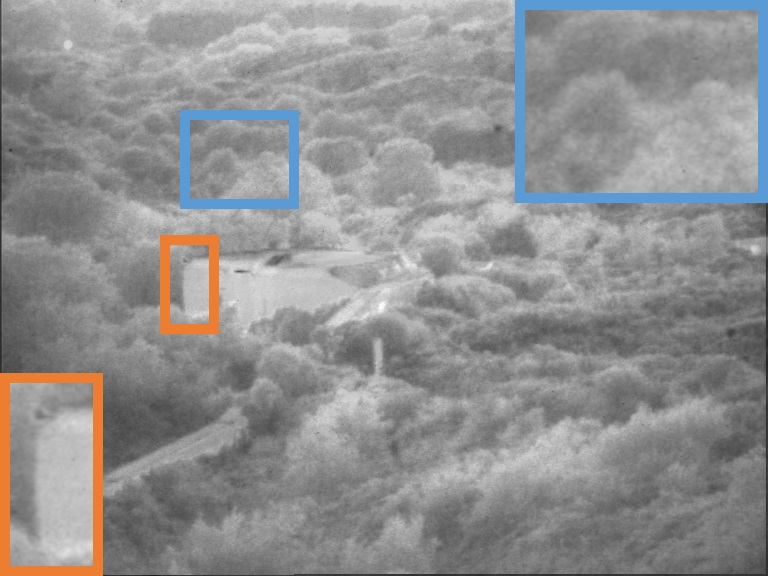}} 
	\subfigure[FusionGan]{\includegraphics[width=0.16\textwidth]{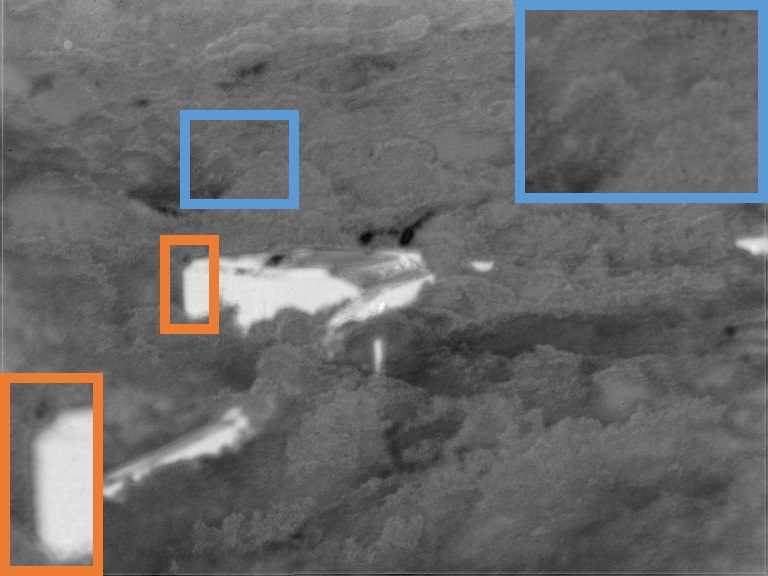}} 
	\subfigure[SDF]{\includegraphics[width=0.16\textwidth]{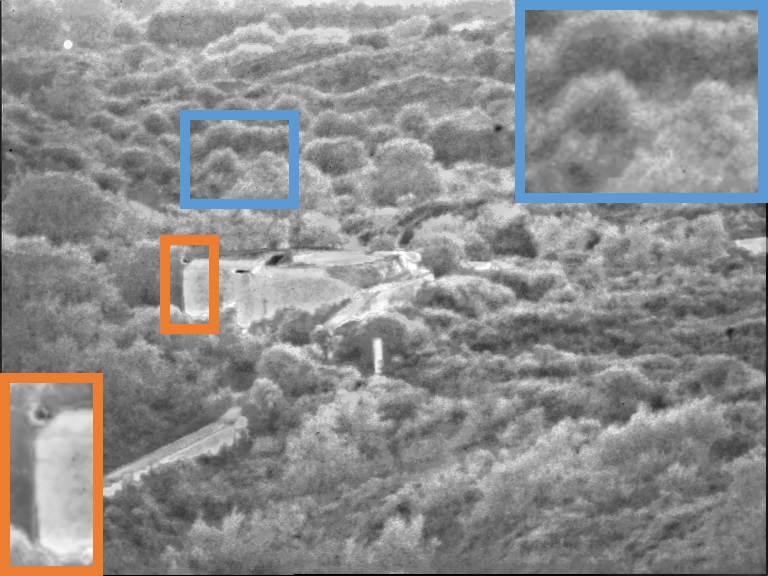}} 
	\subfigure[TVAL]{\includegraphics[width=0.16\textwidth]{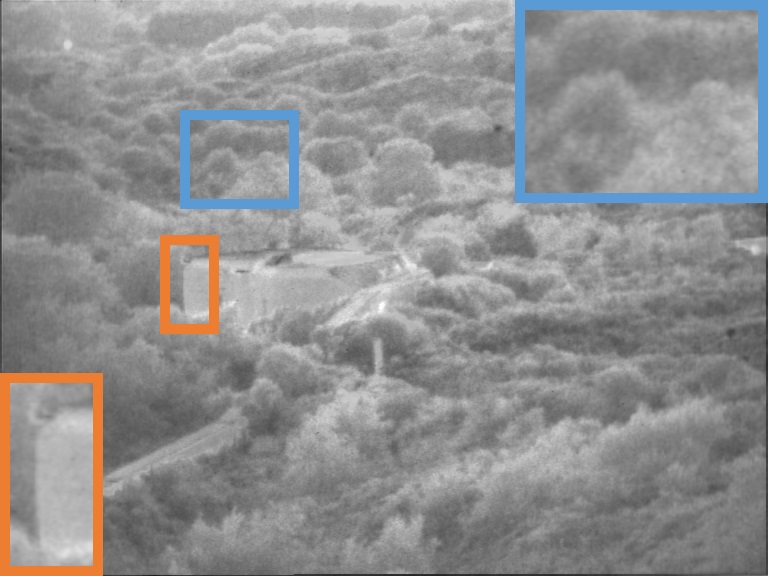}} 
	\subfigure[Ours]{\includegraphics[width=0.16\textwidth]{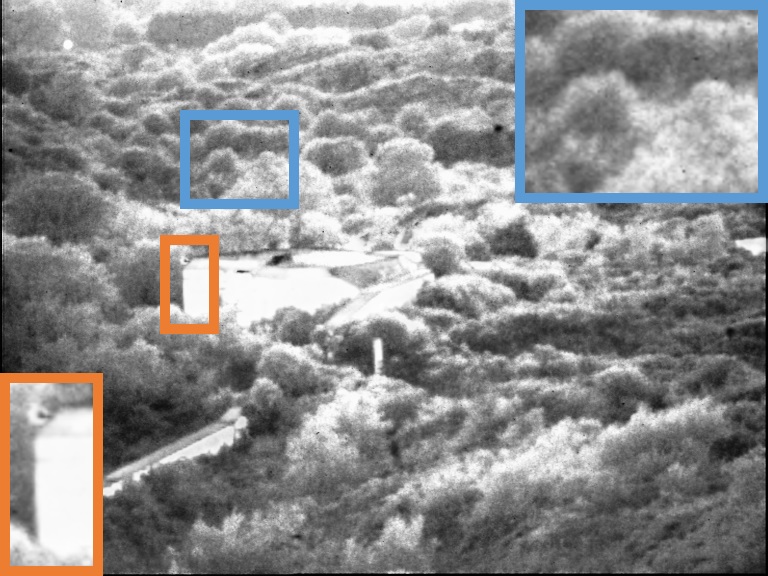}} 
	\caption{The fused images of \textit{Bunker}.}
	\label{fig:Qualitative}
\end{figure*}
\subsubsection{Datasets, Metrics and Details}
As shown in Table~\ref{tab:dataset}, IVF experiments use three datasets (FLIR, NIR and TNO). The 180 pairs of images in FLIR compose the training set. Two subsets (Water and OldBuilding) of NIR are used for validation. To comprehensively evaluate the performance of different models, we employ TNO, NIR-Country and the rest pairs of FLIR as test datasets. To the best of our knowledge, most of the papers only employ part of cherry-picked pairs in TNO as test sets. However, our test sets contain more than 130 pairs with different illuminations and scenarios. To quantitatively measure the fusion performance, six metrics are employed: entropy~(EN)~\cite{roberts2008assessment}, standard deviation~(SD)~\cite{rao1997fibre}, spatial frequency~(SF)~\cite{Eskicioglu19952959}, visual information fidelity~(VIF)~\cite{HAN2013127}, average gradient~(AG)~\cite{CUI2015199} and sum of the correlations of differences~(SCD)~\cite{aslantas2015new}. Larger metrics indicate that a fusion image is better. In our experiment, the tuning parameter $\lambda^{\rm IVF}$ in Eq.~(\ref{equ:Loss_total}) is set to 5. The network is optimized over 60 epochs with a learning rate of $10^{-2}$ in the first 30 epochs and $10^{-3}$ in the rest epochs. The number of DCUs, activation function and fusion strategy may significantly affect the performance of CSC-IVFN. We determine them on validation sets. With the limited space, the validation experiments are exhibited in supplementary materials and the best configuration is reported as follows: the number of DCUs in base or detail encoder is 7; the activation functions in base and detail encoders are set as PReLU and SST, respectively; the fusion strategies for base and detail images are saliency-weighted fusion and IVF $\ell_1$-norm fusion, respectively.

\subsubsection{Comparison with SOTA Methods}

To verify the superiority of our CSC-IVFN, we compare its fusion results with nine popular IVIF fusion methods, including
ADKT~\cite{bavirisetti2015fusion},
CSR~\cite{liu2016image},
DeepFuse~\cite{prabhakar2017deepfuse},
DenseFuse~\cite{li2018densefuse},
DLF~\cite{li2018infrared},
FEZL~\cite{DBLP:journals/corr/abs-1905-03590},
FusionGAN~\cite{ma2019fusiongan},
SDF~\cite{bavirisetti2016two} and
TVAL~\cite{guo2017infrared}.
Six metrics of all methods are displayed in Table~\ref{tab:QuantitativeResult}. It is shown that our method achieves the best performance on all test sets with regard to most metrics. Therefore, our method is suitable for various scenarios with different kinds of illuminations and object categories. In contrast, the other methods (including DeepFuse, DenseFuse and SDF) can achieve good performance on certain test sets with regard to a part of metrics. Besides the metric comparison, representative fusion images are displayed in Fig.~\ref{fig:Qualitative}. In the visible image, there are lots of bushes. In the infrared image, we can observe a bunker. However, it is not easy to recognize the bushes/bunker in the infrared/visible image. It is found that our fusion image keeps the details and textures of the visible image, and preserves the interest objects (i.e., the bushes and the bunker). In addition, its contrast is fairly high. In conclusion, both visible spectrum and thermal radiation information are retained in our fusion image. However, other methods cannot generate satisfactory images as good as ours.

\begin{table*}[!]
	\centering
	\caption{Quantitative results of the IVF task. Boldface and underline indicate the best and the second best results, respectively.}
	\resizebox{.9\textwidth}{!}{
	\begin{tabular}{ccccccccccc}
		\toprule
		\multicolumn{11}{c}{\textbf{Dataset: FLIR}}\\
		& ADKT    & CSR     & DeepFuse      & DenseFuse     & DLF     & FEZL    & FusianGAN & SDF           & TVAL    & Ours             \\
		\midrule
		EN          & 6.80  & 6.91  & \underline{7.21}    & 7.21          & 6.99       & 6.91  & 7.02      & 7.15        & 6.80  & \textbf{7.61}  \\
		MI          & 2.72  & 2.57  & 2.73          & 2.73          & \underline{2.78} & 2.78  & 2.68      & 2.31        & 2.47  & \textbf{3.02}  \\
		SD          & 28.37 & 30.53 & \underline{37.35}   & 37.32         & 32.58      & 31.16 & 34.38     & 35.89       & 28.07 & \textbf{55.94} \\
		SF          & 14.48 & 17.13 & 15.47         & 15.50         & 14.52      & 14.16 & 11.51     & \underline{18.79} & 14.04 & \textbf{21.85} \\
		VIF         & 0.34  & 0.37  & 0.50          & 0.50          & 0.42       & 0.33  & 0.29      & \underline{0.50}  & 0.33  & \textbf{0.70}  \\
		AG          & 3.56  & 4.80  & 4.80          & 4.82          & 4.15       & 3.38  & 3.20      & \underline{5.57}  & 3.52  & \textbf{6.92}  \\
		SCD         & 1.39  & 1.42  & 1.72          & \underline{1.72}    & 1.57       & 1.42  & 1.18      & 1.50        & 1.40  & \textbf{1.80}  \\
		\midrule
		\multicolumn{11}{c}{\textbf{Dataset: NIR-Country Scene}}\\
		& ADKT    & CSR     & DeepFuse      & DenseFuse     & DLF     & FEZL    & FusianGAN & SDF           & TVAL    & Ours             \\
		\midrule
		EN          & 7.11  & 7.17  & 7.30          & \underline{7.30}    & 7.22       & 7.19  & 7.06      & 7.30        & 7.13  & \textbf{7.36}  \\
		MI          & 3.94  & 3.70  & \underline{4.04}    & \textbf{4.04} & 3.97       & 3.81  & 3.00      & 3.29        & 3.67  & 3.86           \\
		SD          & 38.98 & 40.38 & 45.82         & \underline{45.85}   & 42.31      & 44.44 & 34.91     & 43.74       & 40.47 & \textbf{69.37} \\
		SF          & 17.31 & 20.37 & 18.63         & 18.72         & 18.36      & 17.04 & 14.31     & \underline{20.65} & 16.69 & \textbf{28.29} \\
		VIF         & 0.54  & 0.58  & 0.68          & 0.68          & 0.61       & 0.55  & 0.42      & \underline{0.69}  & 0.53  & \textbf{1.05}  \\
		AG          & 5.38  & 6.49  & 6.18          & 6.23          & 5.92       & 5.38  & 4.56      & \underline{6.82}  & 5.32  & \textbf{9.42}  \\
		SCD         & 1.09  & 1.12  & 1.37          & \underline{1.37}    & 1.22       & 1.14  & 0.51      & 1.19        & 1.09  & \textbf{1.73}  \\
		\midrule
		\multicolumn{11}{c}{\textbf{Dataset: TNO}}\\
		& ADKT    & CSR     & DeepFuse      & DenseFuse     & DLF     & FEZL    & FusianGAN & SDF           & TVAL    & Ours             \\
		\midrule
		EN          & 6.40  & 6.43  & \underline{6.86}    & 6.84          & 6.38       & 6.63  & 6.58      & 6.67        & 6.40  & \textbf{6.91}  \\
		MI          & 2.01  & 1.99  & 2.30          & 2.30          & 2.15       & 2.23  & \underline{2.34}& 1.72        & 2.04  & \textbf{2.50}  \\
		SD          & 22.96 & 23.60 & \underline{32.25}   & 31.82         & 22.94      & 28.05 & 29.04     & 28.04       & 23.01 & \textbf{46.97} \\
		SF          & 10.78 & 11.44 & 11.13         & 11.09         & 9.80       & 9.46  & 8.76      & \underline{12.60} & 9.03  & \textbf{12.88} \\
		VIF         & 0.29  & 0.31  & \underline{0.58}    & 0.57          & 0.31       & 0.31  & 0.26      & 0.46        & 0.28  & \textbf{0.62}  \\
		AG          & 2.99  & 3.37  & 3.60          & 3.60          & 2.72       & 2.55  & 2.42      & \underline{3.98}  & 2.52  & \textbf{4.22}  \\
		SCD         & 1.61  & 1.63  & \textbf{1.80} & \underline{1.80}    & 1.62       & 1.67  & 1.40      & 1.68        & 1.60  & 1.70\\
		\bottomrule
	\end{tabular}
}
	\label{tab:QuantitativeResult}
\end{table*}

\subsection{Multi-Exposure Image Fusion}
\subsubsection{Datasets, Metrics and Details}
Three datasets SICE \cite{SICE}, TCI2018 \cite{MEF-SSIM2018TCI} and HDRPS \footnote{\url{http://markfairchild.org/HDR.html}} are employed in our experiments. HDRPS and TCI2018 are used for test and validation, respectively. SICE is a large and high-quality dataset. It is divided into two parts for training and validation. The basic information of datasets is shown in Table \ref{tab:dataset}. Many papers use MEFSSIM to evaluate the performance, but CSC-MEFN is supervised by MEFSSIM. Hence, it is unfair for other methods. As an alternative, we utilize four SOTA blind image quality indices , i.e., blind/referenceless image spatial quality evaluator (Brisque) \cite{brisque}, naturalness image quality evaluator (Niqe) \cite{Niqe}, perception based image quality evaluator (Piqe) \cite{Piqe} and multi-task end-to-end optimized deep neural network (MEON) based blind image quality assessment \cite{MEON}. Smaller values indicate that a fusion image is better. Experiments show that large $\lambda^{\rm MEF}$ makes training unstable, so at the $i$th iteration it is set to $\min\{0.25(i-1),\lambda^{\rm MEF}_{\max}\}$. We select $\lambda^{\rm MEF}_{\max}=10$ to make halo loss and MEFSSIM loss have similar magnitudes. The network is optimized by Adam over 50 epochs with a learning rate of $5\times10^{-4}$. The network configuration is determined by validation datasets. We utilize $N=3$ DCUs to extract codes and SST is employed as an activation function. 

\subsubsection{Comparison with SOTA Methods}
\begin{figure*}[!htb]
	\centering
	\subfigure[EF]{\includegraphics[width=0.24\textwidth]{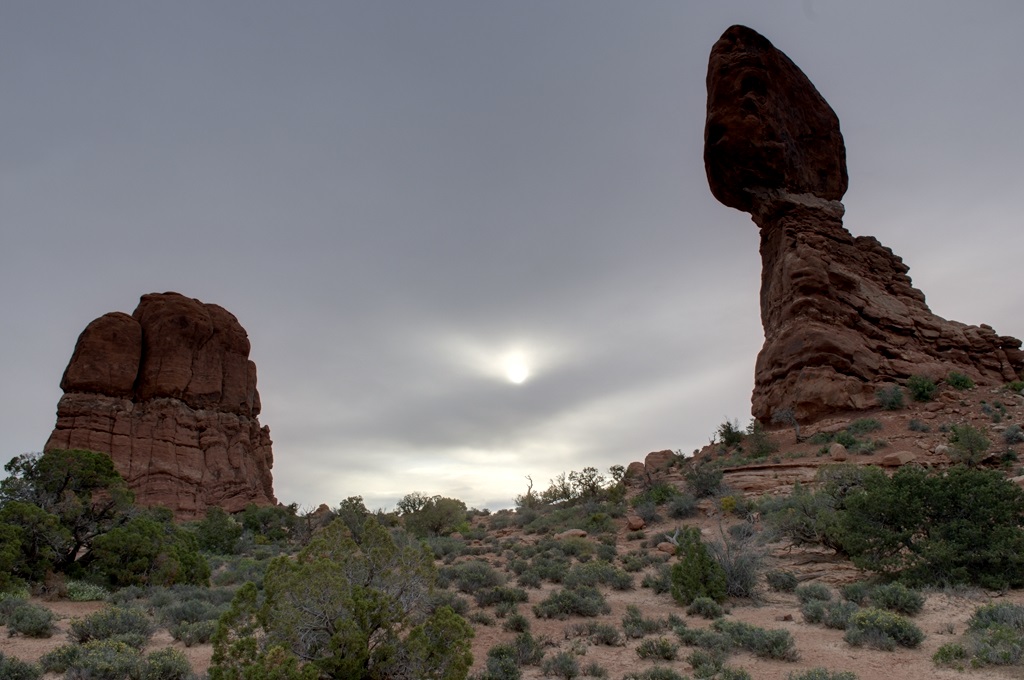}} 
	\subfigure[GGIF]{\includegraphics[width=0.24\textwidth]{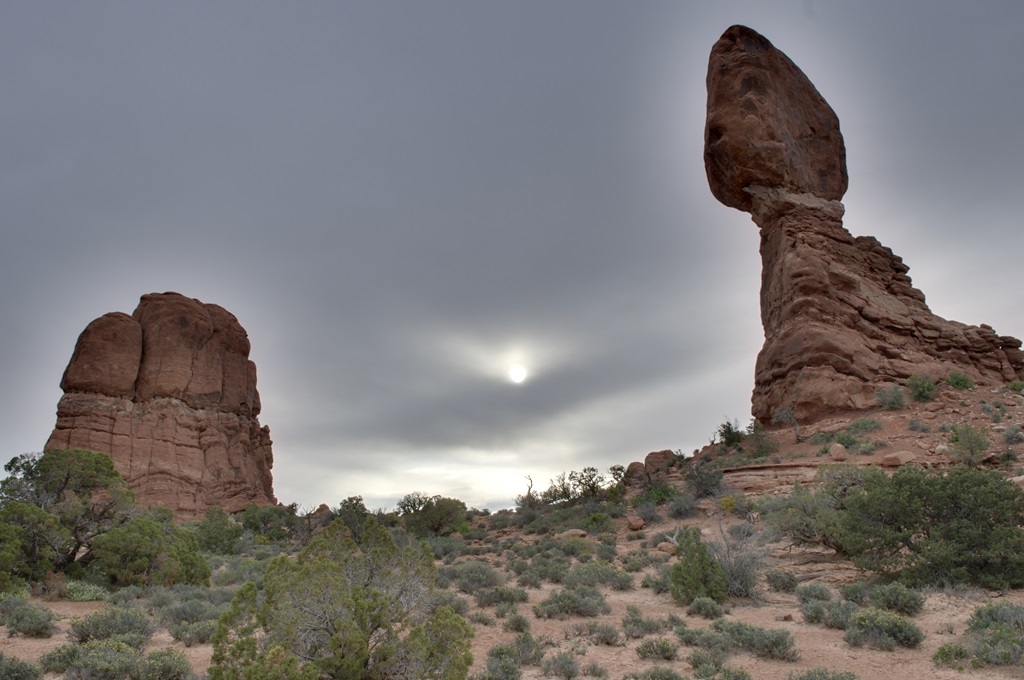}}
	\subfigure[DenseFuse]{\includegraphics[width=0.24\textwidth]{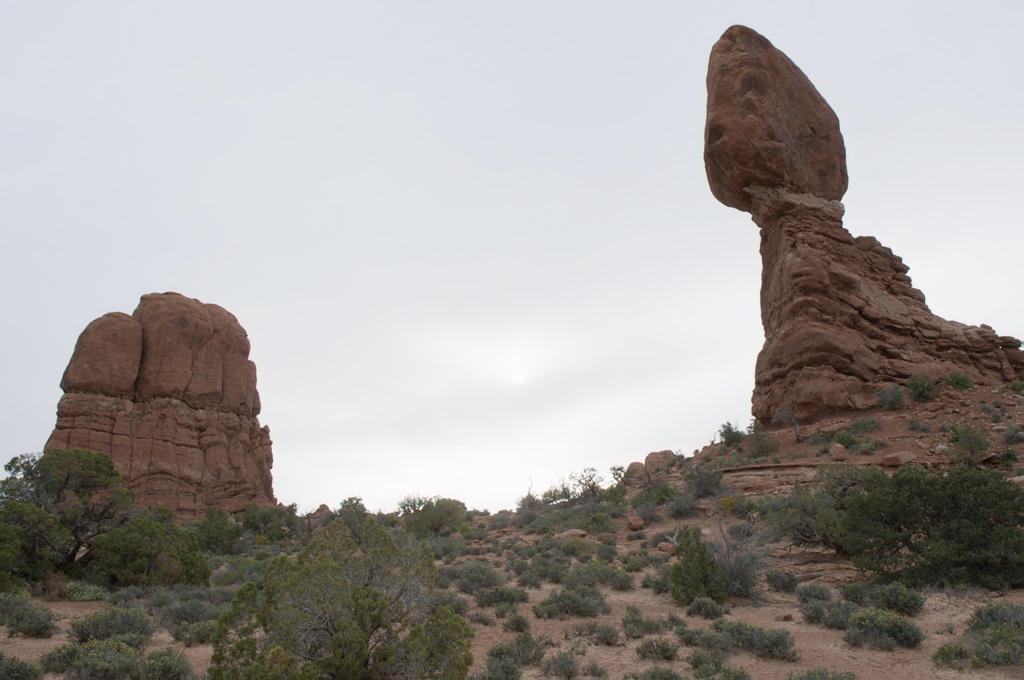}} 
	\subfigure[MEF-Net]{\includegraphics[width=0.24\textwidth]{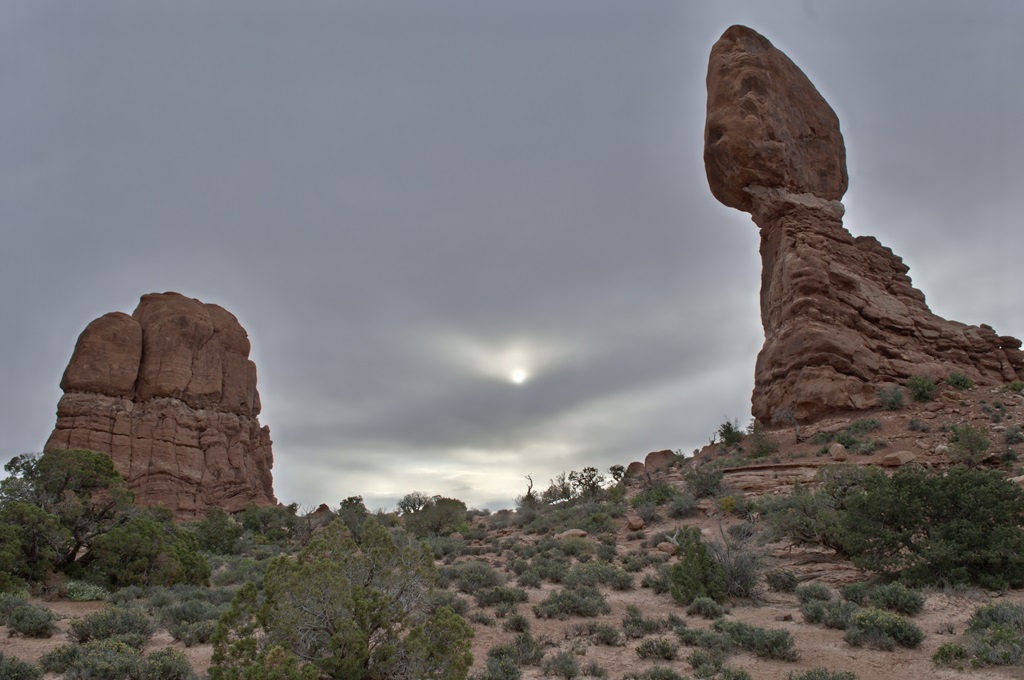}} 
	\subfigure[FMMR]{\includegraphics[width=0.24\textwidth]{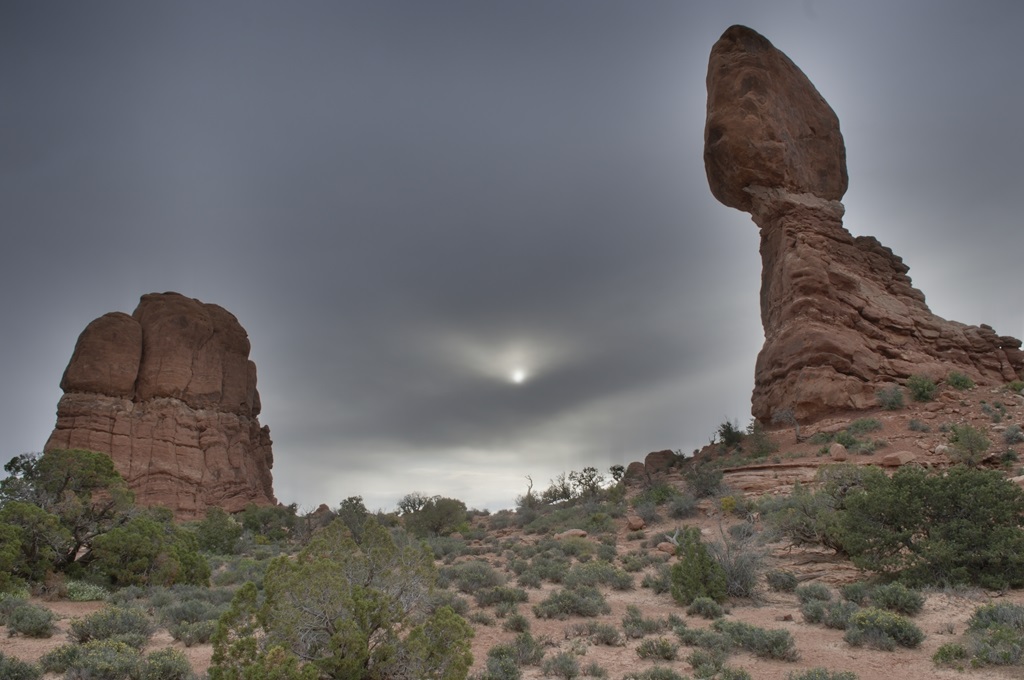}} 
	\subfigure[DSIFTEF]{\includegraphics[width=0.24\textwidth]{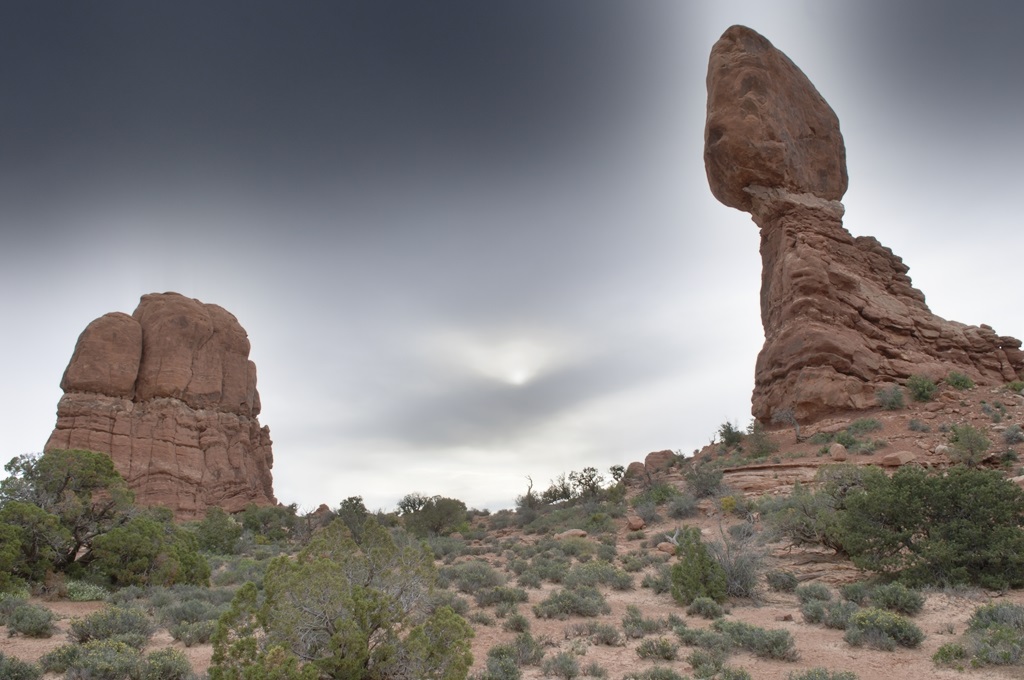}} 
	\subfigure[Lee18]{\includegraphics[width=0.24\textwidth]{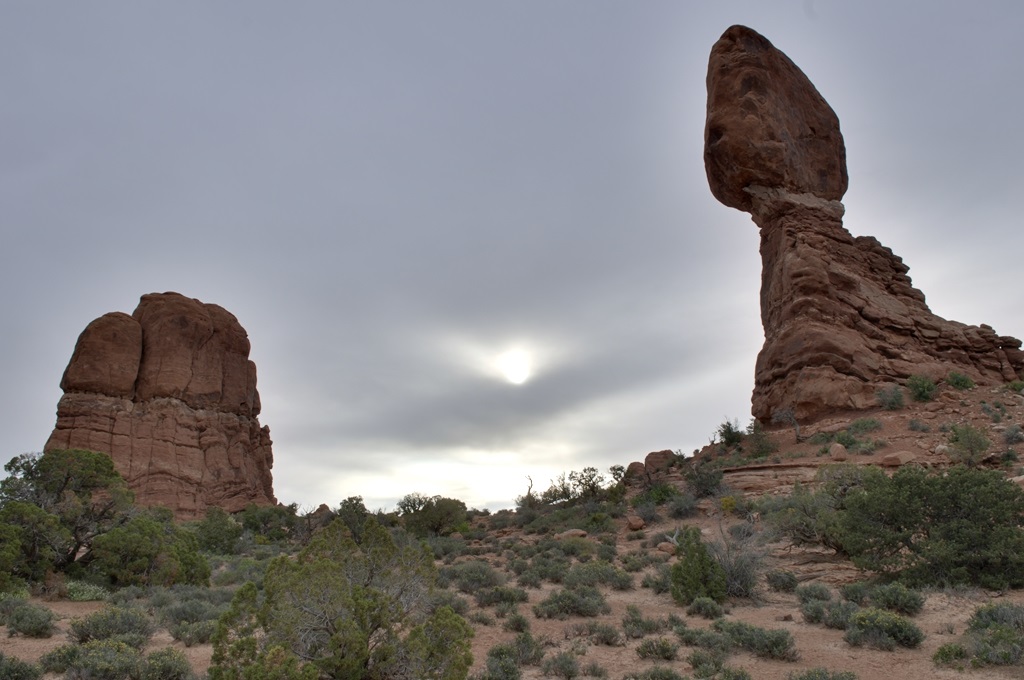}} 
	\subfigure[Ours]{\includegraphics[width=0.24\textwidth]{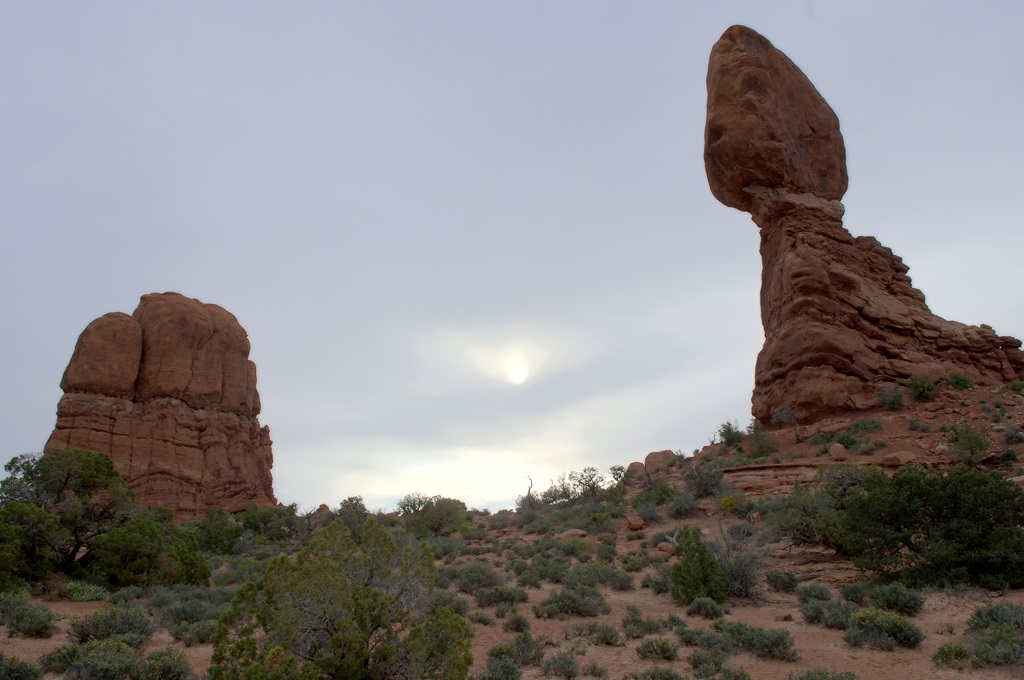}} 
	\caption{The fused images of \textit{Balanced Rock}.}
	\label{fig:MEF}
\end{figure*}

CSC-MEFN is compared with seven classic and recent SOTA methods, including EF \cite{EF},
GGIF \cite{GGIF},
DenseFuse \cite{li2018densefuse},
MEF-Net \cite{MEF-NET},
FMMR \cite{FMMR},
DSIFTEF \cite{DSIFT_EF},
Lee18 \cite{Lee18}.
The metrics are listed in Table \ref{tab:MEF_Metric}. Our network outperforms other methods. Lee18 and EF are ranked in the second and third places. Fig. \ref{fig:MEF} displays the fusion images. It is shown that GGIF, MEF-Net, FMMR, DSIFTEF and Lee18 suffer from strongly halo effects around edges between the sky and rocks. For EF the right rock is too dark, and for DenseFuse the sun cannot be recognized. The contrast of local regions for both EF and DenseFuse is low. Our fusion image strikes the balance. 

\begin{table*}[htbp]
	\centering
	\caption{Quantitative results of the MEF task. Boldface and underline indicate the best and the second best results, respectively.}
	\resizebox{.9\linewidth}{!}{
\begin{tabular}{ccccccccc}
	\toprule
	& EF    & GGIF  & DenseFuse & MEF-Net & FMMR  & DSIFTEF & Lee18 & Ours \\
	\midrule
	MEON  & \underline{8.6730} & 9.1537  & 11.8453  & 9.3623  & 9.8616  & 9.3787  & 9.8093  & \textbf{8.1776} \\
	Brisque & 18.8259  & 19.1711  & 26.4427  & 19.4511  & 20.1099  & 18.6533  & \underline{18.5110} & \textbf{18.2694} \\
	Niqe  & 2.9086  & 2.5204  & 2.5772  & 2.5215  & 2.5494  & 2.5277  & \underline{2.4655} & \textbf{2.3980} \\
	Piqe  & 31.0617  & 32.1874  & \underline{29.6126} & 32.2904  & 32.0856  & 32.2915  & 32.5380  & \textbf{27.8342} \\
	\bottomrule
\end{tabular}%
	}
	\label{tab:MEF_Metric}%
\end{table*}%

\subsection{Multi-Modal Image Fusion}
\subsubsection{Datasets, Metrics and Details}
As shown in Table \ref{tab:dataset}, we employ a multispectral/RGB image fusion dataset, Cave \cite{CAVE_0293}. It contains 32 scenes, each of which has a 31-band multispectral image and an RGB image. It is divided into three parts for training, testing and validation. The Wald protocol is used to construct training sets. We employ peak signal-to-noise ratio (PSNR) and SSIM as evaluation indexes. Larger PSNR and SSIM indicate that a fusion image is better. The network is optimized by Adam over 100 epochs with a learning rate of $5\times10^{-4}$. SST is employed as an activation function. The number of DCUs is empirically set to 4 for a speed and accuracy trade-off. 

\subsubsection{Comparison with SOTA Methods}
\begin{figure*}[!htb]
	\centering
	\subfigure[CNMF]{\includegraphics[width=0.24\textwidth]{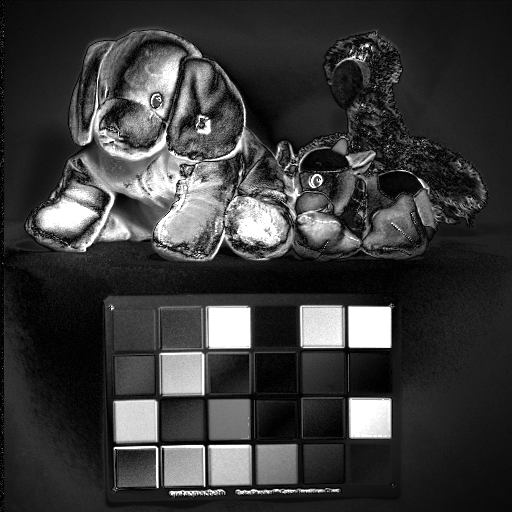}} 
	\subfigure[GSA]{\includegraphics[width=0.24\textwidth]{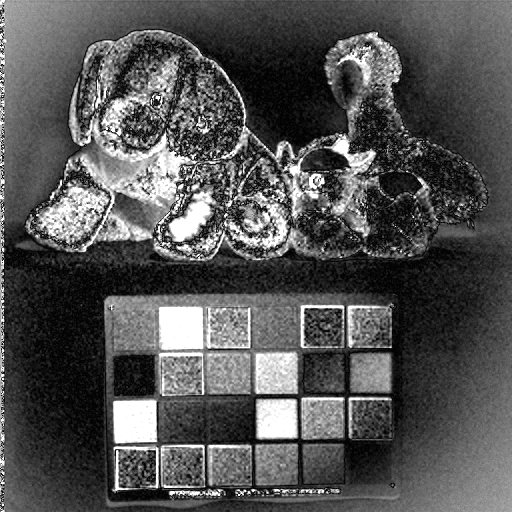}}
	\subfigure[FUSE]{\includegraphics[width=0.24\textwidth]{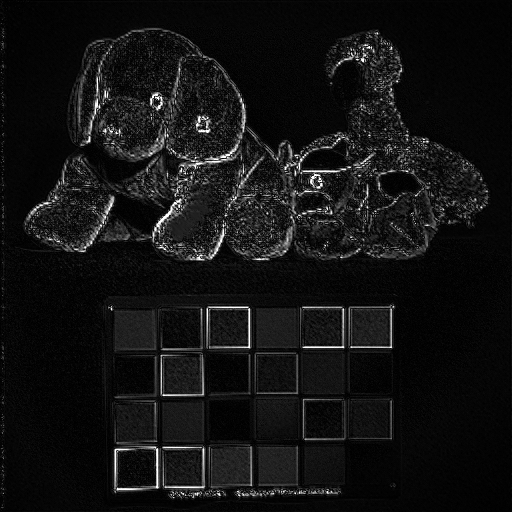}} 
	\subfigure[MAPSMM]{\includegraphics[width=0.24\textwidth]{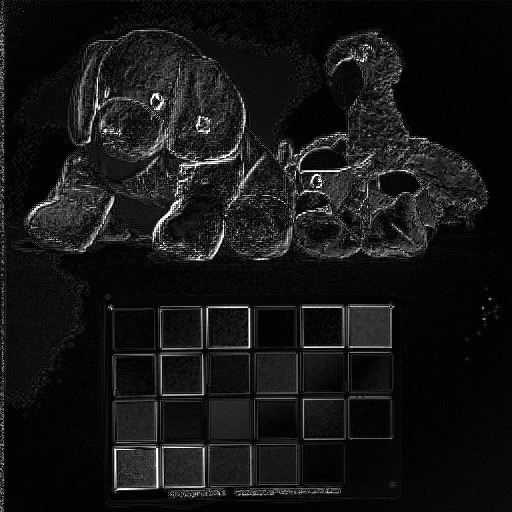}} 
	\subfigure[GLPHS]{\includegraphics[width=0.24\textwidth]{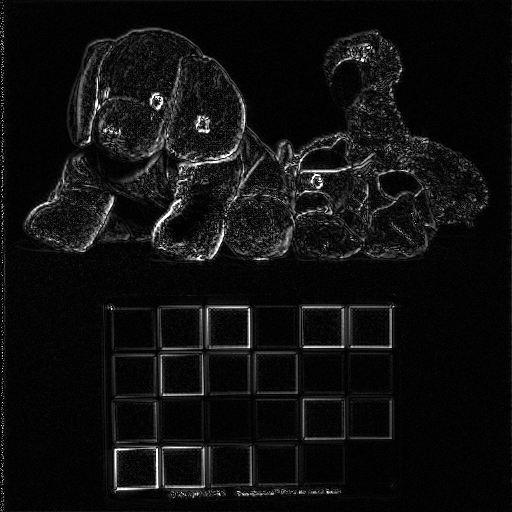}} 
	\subfigure[PNN]{\includegraphics[width=0.24\textwidth]{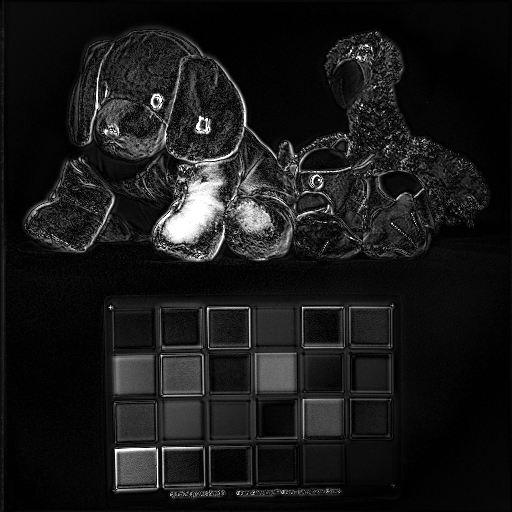}} 
	\subfigure[PFCN]{\includegraphics[width=0.24\textwidth]{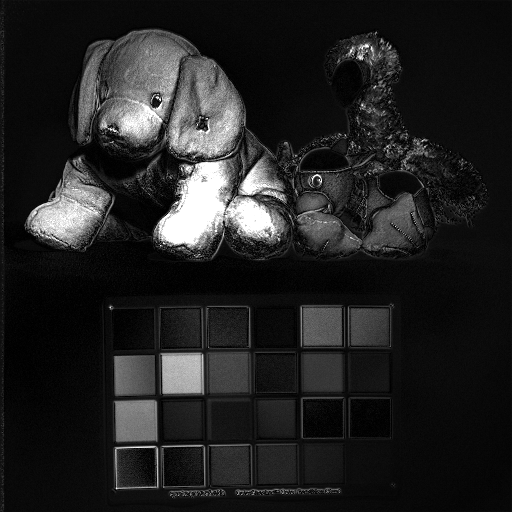}} 
	\subfigure[Ours]{\includegraphics[width=0.24\textwidth]{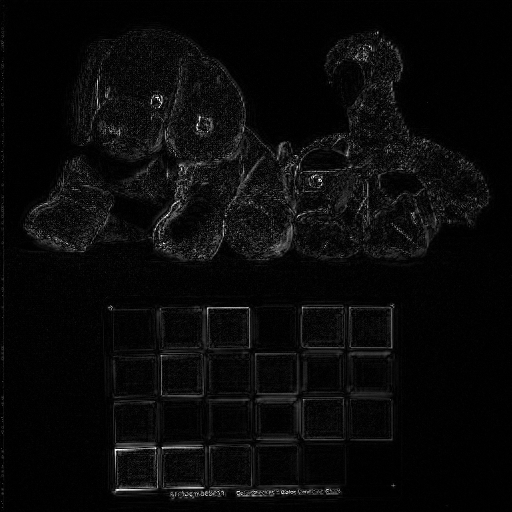}} 
	\caption{The error maps of \textit{stuffed toys} (band 3). Their values are amplified 10 times for easier visual inspection. The error goes larger from black to white.}
	\label{fig:MMF}
\end{figure*}

CSC-MMFN is compared with seven classic and recent SOTA methods, including CNMF \cite{CNMF}, GSA \cite{GSA}, FUSE \cite{FUSE}, MAPSMM \cite{MAPSMM}, GLPHS \cite{GLP}, PNN \cite{PNN} and PFCN \cite{PFCN}. The metrics listed in Table \ref{tab:MMF} show that our network achieves the largest PSNR and SSIM. GLPHS and PFCN can be ranked in the second place in terms of PSNR and SSIM, respectively. The error maps of the third band of \textit{stuffed toys} are displayed in Fig. \ref{fig:MMF}. We found that CNMF, GSA and PFCN break down when reconstructing the color checkerboard and stuffed toys, while FUSE, MAPSMM, GLPHS and PNN perform badly at the edges. In summary, CSC-MMFN has the best performance.

\begin{table*}[htbp]
	\centering
	\caption{Quantitative results of the MMF task. Boldface and underline indicate the best and the second best results, respectively.}
\resizebox{.9\textwidth}{!}{
\begin{tabular}{lccccccccrcc}
	\toprule
	\multicolumn{1}{c}{\multirow{2}[4]{*}{Images}} & \multicolumn{2}{c}{CNMF} &       & \multicolumn{2}{c}{GSA} &       & \multicolumn{2}{c}{FUSE} &       & \multicolumn{2}{c}{MAPSMM} \\
	\cmidrule{2-3}\cmidrule{5-6}\cmidrule{8-9}\cmidrule{11-12}      & PSNR  & SSIM  &       & PSNR  & SSIM  &       & PSNR  & SSIM  &       & PSNR  & SSIM \\
	\midrule
	R\&F apples & 34.5743  & 0.9384  &       & 32.7312  & 0.6816  &       & 38.2509  & 0.9434  &       & 41.4403  & 0.9786  \\
	R\&F peppers & 33.1338  & 0.9305  &       & 30.9636  & 0.7026  &       & 35.7674  & 0.9177  &       & 39.5621  & 0.9670  \\
	Sponges & 31.1378  & 0.9549  &       & 26.3144  & 0.7429  &       & 33.7565  & 0.9368  &       & 35.2542  & 0.9347  \\
	Stuffed toys & 30.0417  & 0.8652  &       & 27.3283  & 0.5764  &       & 34.3008  & 0.9372  &       & 36.4635  & 0.9449  \\
	Superballs & 21.2880  & 0.8292  &       & 32.5318  & 0.7626  &       & 36.3646  & 0.9078  &       & 27.5589  & 0.6020  \\
	Thread spools & 32.3698  & 0.8921  &       & 30.6611  & 0.6591  &       & 33.9568  & 0.9088  &       & 34.9208  & 0.9397  \\
	Mean  & 30.4242  & 0.9017  &       & 30.0884  & 0.6875  &       & 35.3995  & 0.9253  &       & 35.8666  & 0.8945  \\
	\midrule
	\multicolumn{1}{c}{\multirow{2}[4]{*}{Images}} & \multicolumn{2}{c}{GLPHS} &       & \multicolumn{2}{c}{PNN} &       & \multicolumn{2}{c}{PFCN} &       & \multicolumn{2}{c}{Ours} \\
	\cmidrule{2-3}\cmidrule{5-6}\cmidrule{8-9}\cmidrule{11-12}      & PSNR  & SSIM  &       & PSNR  & SSIM  &       & PSNR  & SSIM  &       & PSNR  & SSIM \\
	\midrule
	R\&F apples & \underline{43.5554} & \underline{0.9873} &       & 39.9322  & 0.9681  &       & 41.5981  & 0.9864  &       & \textbf{51.5897} & \textbf{0.9954} \\
	R\&F peppers & \underline{41.6063} & 0.9822  &       & 39.4820  & 0.9666  &       & 40.4695  & \underline{0.9835} &       & \textbf{49.5495} & \textbf{0.9947} \\
	Sponges & \underline{37.2994} & 0.9735  &       & 31.3927  & 0.9573  &       & 32.0306  & \underline{0.9830} &       & \textbf{43.2901} & \textbf{0.9873} \\
	Stuffed toys & \underline{38.3917} & \underline{0.9756} &       & 33.6743  & 0.9585  &       & 33.1012  & 0.9713  &       & \textbf{44.1118} & \textbf{0.9897} \\
	Superballs & \underline{39.3176} & 0.9494  &       & 36.9901  & 0.9533  &       & 36.7382  & \underline{0.9756} &       & \textbf{46.2919} & \textbf{0.9873} \\
	Thread spools & 36.3586  & 0.9558  &       & 35.8109  & 0.9540  &       & \underline{38.8272} & \underline{0.9863} &       & \textbf{42.5585} & \textbf{0.9857} \\
	Mean  & \underline{39.4215} & 0.9706  &       & 36.2137  & 0.9596  &       & 37.1275  & \underline{0.9810} &       & \textbf{46.2319} & \textbf{0.9900} \\
	\bottomrule
\end{tabular}%
}
	\label{tab:MMF}%
\end{table*}%

\section{Conclusion}\label{sec:conclusion}
Inspired by converting the ISTA and CSC models into a hidden layer of neural networks, this paper proposes three deep CSC networks for IVF, MEF and MMF tasks. Extensive experiments and comprehensive comparisons demonstrate that our networks outperform the SOTA methods. Furthermore, the experiments in supplementary materials show that our networks are highly reproducible.

{\small
	\bibliographystyle{plain}
	\bibliography{egbib}
}

\end{document}